\documentclass[lettersize,journal]{IEEEtran}
\usepackage{amsmath,amsfonts}
\usepackage{algorithmic}
\usepackage{algorithm}
\usepackage{array}
\usepackage[caption=false,font=normalsize,labelfont=sf,textfont=sf]{subfig}
\usepackage{textcomp}
\usepackage{stfloats}
\usepackage{url}
\usepackage{verbatim}
\usepackage{graphicx}
\usepackage{cite}
\usepackage{bm} 
\usepackage{gensymb}
\usepackage{hyperref}
\usepackage{inputenc}
\usepackage{amsmath}
\usepackage{xcolor}
\usepackage{soul}
\DeclareMathAlphabet\mathrsfso      {U}{rsfso}{m}{n}
\usepackage{xcolor}

\begin{document}
\title{Hybrid Fingerprint-based Positioning \\in Cell-Free Massive MIMO Systems}
\author{Manish Kumar, Tzu-Hsuan Chou, Byunghyun Lee, Nicolò Michelusi, David J. Love and James V. Krogmeier
\thanks{Manish Kumar, Byunghyun Lee, D. J. Love, and J. V. Krogmeier are with the School of Electrical and Computer Engineering, Purdue University, West Lafayette, IN, USA; emails: \{mkrishne, lee4093, djlove, jvk\}@purdue.edu.}
\thanks{T.-H. Chou is with Qualcomm, Inc., San Diego, CA 92121 USA; e-mail: tzuhchou@qti.qualcomm.com.}
\thanks{N. Michelusi is with the School of Electrical, Computer and Energy Engineering, Arizona State University, AZ, USA; email: nicolo.michelusi@asu.edu. His research is funded by NSF CNS-2129015.}
\thanks{This work is supported in part by the National Science Foundation under grants EEC-1941529, CNS-2235134, CNS-2212565, CNS-2225578 and the Office of Naval Research under grant N000142112472.}
\vspace{-8mm}
}
\maketitle
\begin{abstract}
Recently, there has been an increasing interest in 6G technology for integrated sensing and communications, where positioning stands out as a key application. In the realm of 6G, cell-free massive multiple-input multiple-output systems, featuring distributed base stations equipped with a large number of antennas, present an abundant source of angle-of-arrival (AOA) information that could be exploited for positioning applications. In this paper we leverage this AOA information at the base stations using the multiple signal classification algorithm, in conjunction with received signal strength (RSS) for positioning through Gaussian process regression (GPR). An AOA fingerprint database is constructed by capturing the angle data from multiple locations across the network area and is combined with RSS data from the same locations to form a hybrid fingerprint which is then used to train a GPR model employing a squared exponential kernel. The trained regression model is subsequently utilized to estimate the location of a user equipment. Simulations demonstrate that the GPR model with hybrid input achieves better positioning accuracy than traditional GPR models utilizing RSS-only and AOA-only inputs.  
\end{abstract}


\vspace{-3.5mm}
\section{Introduction}
\IEEEPARstart{G}{eospatial} positioning and localization technologies play an essential role in providing users with accurate and contextually relevant information. As cellular wireless communication evolves, the advent of distributed and cell-free massive multiple-input multiple-output (MIMO) systems, regarded as a potential cornerstone for 6G, offers promising research opportunities in the domains of positioning and localization~\cite{Love_6G_paper}. Recent studies have shown that among various positioning techniques, fingerprint-based positioning is particularly effective in these systems \cite{fingerprint_savic,joint_aoa_rss,2D_fingerprint_mmwave}. It benefits from the wealth of position-related data provided by multiple base stations and is largely unaffected by the non-line-of-sight~(NLOS) bias induced by multipath propagation.

A fingerprint in the context of positioning technology is a data representation that captures key information about the radio signal characteristics in a given en zvironment. Such fingerprints typically consist of measurements such as received signal strength (RSS), angle-of-arrival (AOA) and channel impulse response estimates. These measurements can be systematically collected at predefined locations within the environment to construct a comprehensive fingerprint database. When similar measurements are obtained from a device at an unknown location, they are compared with the stored fingerprint database using various positioning algorithms. Following this
comparative analysis, the unknown location is estimated.

Numerous fingerprint-based positioning algorithms have been widely investigated for indoor environments \cite{wifi_overview}. However, research focusing specifically on the application of these algorithms to distributed and cell-free massive MIMO systems is still limited. Some of the notable works in distributed MIMO systems employ techniques such as linear regression (LR), weighted k-nearest neighbors (WKNN), neural networks \cite{fog_massive_mimo_ml}, and Gaussian process regression (GPR) \cite{fingerprint_savic,ml_rss_surya} with RSS fingerprints. Despite its computational overhead, GPR is widely used in the research community due to its ability to provide probabilistic outputs and has shown superior positioning accuracy compared to other techniques \cite{ml_rss_surya}. However, most existing studies on GPR primarily rely on RSS inputs, with limited investigation into other fingerprint types.

AOA-based fingerprints are relatively more prevalent than RSS fingerprints in positioning research for massive MIMO systems, owing to the greater consistency of AOA measurements. Recent studies have investigated the application of AOA-based fingerprints in cell-free massive MIMO systems, using techniques such as k-means clustering \cite{joint_aoa_rss,sec_trans_positioning} and maximum likelihood estimation \cite{2D_fingerprint_mmwave}. In particular, \cite{joint_aoa_rss} shows that combining AOA and RSS measurements through a joint angle-domain channel power matrix and RSS fingerprint, leveraging unsupervised k-means clustering and WKNN, outperforms individual fingerprints in positioning accuracy. These studies provide a compelling rationale for incorporating AOA estimates into supervised GPR models. 

In this work, we propose an innovative fingerprint-based positioning approach for cell-free massive MIMO systems, utilizing GPR as the core positioning algorithm. To enhance positioning accuracy relative to traditional GPR models, we incorporate a hybrid of RSS and AOA inputs. Both AOA and RSS measurements are collected from predefined locations to construct a hybrid fingerprint database that integrates both types of information. This hybrid fingerprint, along with location information, serves as input for training a GPR model. The trained GPR model is then used to predict the position of a user at an unknown location, utilizing both RSS measurements and AOA measurements derived from the multiple signal classification (MUSIC) algorithm. 

To the best of our knowledge, this is the first study to explore the incorporation of AOA inputs into a GPR-based positioning framework. Furthermore, we take a step beyond by investigating hybrid fingerprints, which is also novel in the context of GPR-based positioning. The simulation results highlight the effectiveness of the proposed hybrid fingerprinting approach compared to the baseline RSS-based GPR model, as well as WKNN and LR methods~\cite{fog_massive_mimo_ml}.

\vspace{-1mm}
\section{System Model}
\label{system_model}
\begin{figure}
\centering
\captionsetup{justification=centering}
\includegraphics[width=2.5in]{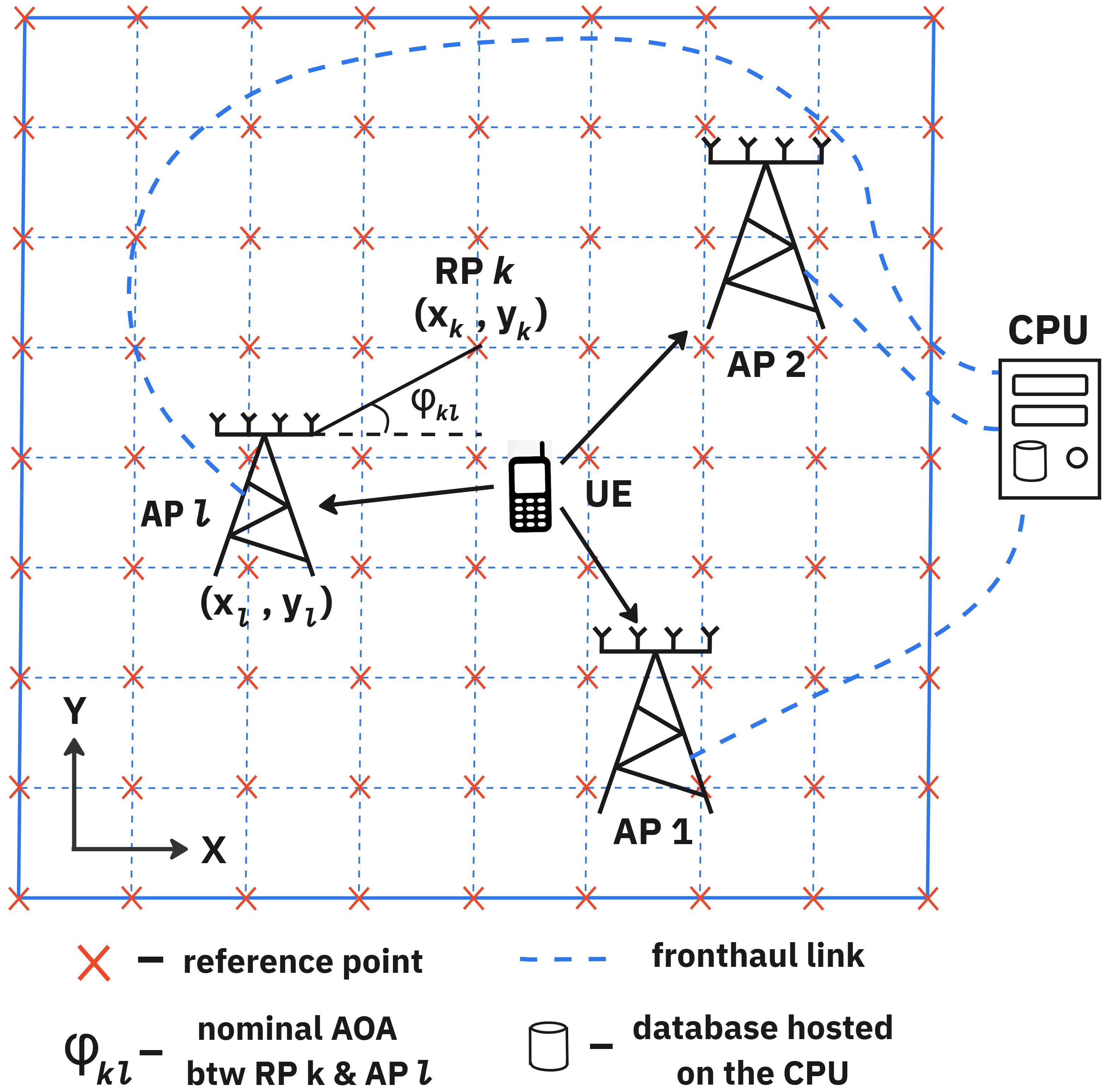}
\caption{System model illustrating the positioning scenario}
\label{system_model_fig}
\vspace{-5mm}
\end{figure}

A cell-free massive MIMO system consisting of $L$ geographically dispersed access points (APs) in the coverage area is considered for our positioning scenario, as shown in Fig. \ref{system_model_fig}. 
Each of the $L$ APs is equipped with a uniform linear array (ULA) of $N$ omnidirectional antennas. Without loss of generality, we assume that the ULA axis is oriented along the x-axis. The APs are connected to an edge-cloud processor called central processing unit (CPU) through fronthaul links. The CPU coordinates the operation of APs within the system.

The positioning process is carried out in two distinct phases: the offline phase and the online phase. In the offline phase, a single-antenna UE is sequentially placed at predefined locations, called reference points (RPs), distributed throughout the system. RSS and AOA measurements from the UE placed at each of the $K$ RPs are collected from all the APs and stored as a fingerprint database on the CPU. The CPU trains a GPR model using RSS and AOA as input features and RP coordinates as target labels. By assuming a Gaussian process prior, the GPR model captures the relationship between inputs and outputs. In the online phase, the trained GPR model predicts the location of the UE using RSS and AOA measurements from a test point, aiming to minimize the prediction error. 

The scatterers around the UE are assumed to be distributed according to a local scattering model\cite{mimo_book_correlation_matrix, disk_scatter_model}. The UE transmits a narrowband pilot vector $\boldsymbol{\psi} \in \mathbb{C}^{z \times 1}$, $\|\boldsymbol{\psi}\|^2 =z$ when placed at an RP during the offline phase or at a test point during the online phase. The pilot vector is assigned to the UE upon network access. In addition, we assume that the UE executes coarse synchronization to correct for rough delay discrepancies in the channel, ensuring that the transmission of the pilot vector $\boldsymbol{\psi}$ is aligned with the network's timing requirements \cite{ofdm_syncr, cell_free_mimo_book}. The received signal $\mathbf{Y}_{k\ell} \in \mathbb{C}^{N \times z } $ at AP $\ell$ from the UE placed at RP $k$ is given by
\begin{equation}
\label{eqn1}
\textbf{Y}_{k\ell} = \sqrt{\rho}\textbf{h}_{k\ell}\boldsymbol{\psi}^H + \textbf{W}_{\ell}.
\end{equation}
Here $\rho$ is the transmit power of UE, $\textbf{W}_{\ell} \in \mathbb{C}^{N \times z}$ denotes the noise at AP $\ell$ with matrix elements independent and identically distributed (i.i.d) as $\mathcal{CN}(0,\sigma_n^2)$ and $\textbf{h}_{k\ell} \in \mathbb{C}^{N \times 1 } $, the channel vector from RP $k$ to AP $\ell$ for a narrowband transmission is given by~\cite{joint_aoa_rss, sec_trans_positioning}
\begin{equation}
\label{eqn2}
\textbf{h}_{k\ell} = \sqrt{\frac{1}{M}}\sum\limits_{m=1}^{M}\sqrt{\beta_{k\ell}}\alpha_{k\ell}^m \textbf{a}(\theta^m_{k\ell}),
\end{equation}
where $M$ denotes the number of scattering paths, which can be arbitrarily large, $\alpha_{k\ell}^m \sim \mathcal{CN}(0,1)$ represents the small-scale fading  coefficient of the $m^{th}$ path, $\theta^m_{k\ell}$ is the azimuth AOA of the $m^{th}$ path and $\mathbf{a}({\theta}^{m}_{k\ell}) \in \mathbb{C}^{N \times 1}$ is the array steering vector given by 
\begin{equation} 
\label{eqn3}
\resizebox{0.9\columnwidth}{!}{$
\textbf{a}(\theta^{m}_{k\ell}) = 
\begin{bmatrix}
1, e^{-j\frac{2\pi d}{\lambda}\cos(\theta^{m}_{k\ell})},\ \dots \ ,e^{-j\frac{2\pi (N-1) d}{\lambda}\cos(\theta^{m}_{k\ell})}
\end{bmatrix}^\top,
$}
\end{equation}
where $d$ is the antenna array spacing and $\lambda$ is the signal carrier wavelength. The angles ${\theta}^{m}_{k\ell}$ are i.i.d random variables modeled as ${\theta}^{m}_{k\ell} = \varphi_{k\ell} + \Theta_m$. Here, $\varphi_{k\ell} = \mathbb{E}[{\theta}^{m}_{k\ell}]$ denotes the nominal azimuth AOA between the UE placed at RP $k$ and AP $\ell$, and corresponds to the azimuth of the straight line connecting the RP and AP. Each i.i.d random variable $\Theta_m$ is characterized by an angular probability density function~(PDF) $f_{\Theta}(\Tilde{\theta})$. This PDF is governed by the underlying scattering model and describes the statistical distribution of angular deviations from the nominal azimuth AOA~\cite{mimo_book_correlation_matrix}. For $\beta_{k\ell}$, the large-scale fading coefficient, a log-distance path loss model is adopted i.e.,
\begin{equation}
\label{eqn4}
    \beta_{k\ell}[dB] = p^{0}_{\ell} - 10\gamma\log_{10}(d_{k\ell}/d_{\ell}^{0}) + \nu_{\scriptscriptstyle k\ell},
\end{equation}
where $d_{k\ell}$ is the three dimensional distance between RP $k$ and the antenna array of AP $\ell$, $p^{0}_{\ell}$ is the path loss (in dB) at reference distance $d^{0}_{\ell}$, $\gamma$ is the path loss exponent  and $\nu_{\scriptscriptstyle k\ell} \sim \mathcal{N}(0,\sigma_{\scriptscriptstyle SF}^2)$  is the shadowing noise. Further, the noise $\textbf{W}_{\ell}$ is assumed to be independent of the transmitted signal $\boldsymbol{\psi}$ and the channel vector $\textbf{h}_{k\ell}$. The APs sample and transmit the received signal to the CPU via fronthaul links for further processing.

\vspace{-1mm}
\section{Fingerprint Extraction and Positioning Strategy}
\subsection{RSS Fingerprint Extraction in the Offline Phase}
Upon receiving the signal $\textbf{Y}_{k\ell}$, the CPU computes the RSS at AP $\ell$ from the UE positioned at RP $k$. The RSS is mathematically expressed as $\xi_{k\ell} = \|\textbf{Y}_{k\ell}\|_F^2$. However, variations in the RSS arise due to small-scale fading in the received signal. To mitigate this, the RSS is calculated as the average value of the signal strength across a sufficiently large number of received symbols, i.e.,~\cite{ml_rss_surya, joint_aoa_rss}
\begin{equation} 
\label{eqn5}
\begin{split}
\widehat{\xi}_{k\ell} &\approx
 \mathbb{E}\{\|\textbf{Y}_{k\ell}\|_F^2\} = \mathbb{E}\{\|{\sqrt{\rho}\textbf{h}_{k\ell}}\boldsymbol{\psi}^H + \textbf{W}_{\ell}\|_F^2\}\\
&= \rho z\mathbb{E}\{\|\textbf{h}_{k\ell}\|^2\} + \mathbb{E}\{\|\textbf{W}_{\ell}\|_F^2\}= \rho zN\beta_{k\ell} + zN\sigma_n^2.
\end{split}
\end{equation} 
The approximation arises from the assumption that by averaging over enough samples, the RSS becomes less sensitive to small-scale fading and additive noise. This effect is a consequence of channel hardening, wherein the small-scale variability of the wireless channel diminishes as the number of independent channel realizations increases, in accordance with the law of large numbers~\cite{cell_free_mimo_book,joint_aoa_rss,ml_rss_surya}. Shadowing, however, persists as it is space-dependent and cannot be averaged out through temporal averaging~\cite{ml_rss_surya,fingerprint_savic}. $\boldsymbol{\xi}_{\ell}~=~[\widehat{\xi}_{\scriptscriptstyle 1\ell}^{\scriptscriptstyle dB}, \widehat{\xi}_{\scriptscriptstyle 2\ell}^{\scriptscriptstyle dB}, \  \dots\  , \widehat{\xi}_{\scriptscriptstyle K\ell}^{\scriptscriptstyle dB}]^\top \in \mathbb{R}^{K\times{1}}$, the vectors of RSS at each AP $\ell$ from all the $K$ RPs, are locally stored in the CPU memory. Here, $\widehat{\xi}_{\scriptscriptstyle k\ell}^{\scriptscriptstyle dB} = 10\log_{10}(\widehat{\xi}_{k\ell}/\rho)$ represents the estimated RSS in decibels~(dB).

\vspace{-1mm}
\subsection{AOA Fingerprint Extraction in the Offline Phase}
The nominal AOA can be determined through geometric measurements during the offline phase using the known locations
of the APs and RPs. For the UE at RP $k$ with coordinates $\boldsymbol{q}_k = (x_k, y_k) \in \mathbb{R}^{1\times2}$ and AP $\ell$  positioned at $(x_{\ell}, y_{\ell})$ as shown in Fig. \ref{system_model_fig}, the nominal azimuth AOA is calculated as $\varphi_{k\ell} = \mathrm{atan2}(y_k - y_\ell, x_k - x_\ell)$, where atan2 denotes the four-quadrant arctangent. The computed AOA vectors $\boldsymbol{\Phi}_{\ell} = [{\varphi}_{\scriptscriptstyle 1\ell}^{\circ}, \varphi_{\scriptscriptstyle 2\ell}^{\circ}, \  \dots\  , \varphi_{\scriptscriptstyle K\ell}^{\circ}]^\top \in \mathbb{R}^{K\times{1}}$ for all $\ell \in \{1,2,\dots,L\}$, along with the RP position matrix $\boldsymbol{Q}_{\scriptscriptstyle RP} = [\boldsymbol{q}_1^\top,\boldsymbol{q}_2^\top,\ \dots \ , \boldsymbol{q}_K^\top]^\top \in \mathbb{R}^{K\times{2}}$ are stored in the CPU memory, where $\varphi_{k\ell}^{\circ}$ represents the AOA measured in degrees. 

\vspace{-1mm}
\subsection{AOA Estimation in the Online Phase}
While RSS estimation in the online phase follows the same procedure as described in the offline phase in (\ref{eqn5}), for estimating the AOA at each AP in the online phase, the CPU employs the MUSIC algorithm~\cite{music_with_scatter}. For a UE located at a test point $\boldsymbol{q}_{\scriptscriptstyle TP} = (x_{\scriptscriptstyle TP},y_{\scriptscriptstyle TP})  \in \mathbb{R}^{1\times2}$, the MUSIC algorithm starts by estimating the covariance matrix of the received signal $\textbf{R}_{\scriptscriptstyle TP,\ell} \in \mathbb{R}^{N \times N}$ at AP $\ell$, expressed as~\cite{mimo_book_correlation_matrix}
\begin{equation} 
\label{eqn6}
\resizebox{0.91\columnwidth}{!}{$
\begin{split}
\textbf{R}_{\scriptscriptstyle TP,\ell} &= \mathbb{E}\{\textbf{Y}_{\scriptscriptstyle TP,\ell}\textbf{Y}_{\scriptscriptstyle TP,\ell}^H\} = \rho z\mathbb{E}\{\textbf{h}_{\scriptscriptstyle TP,\ell}\textbf{h}^H_{\scriptscriptstyle TP,\ell}\} + \mathbb{E}\{\textbf{W}_{\ell}\textbf{W}^H_{\ell}\}\\
&= \rho z\beta_{\scriptscriptstyle TP,\ell}\mathbb{E}\{\textbf{a}({\theta}^m_{\scriptscriptstyle TP,\ell})\textbf{a}^H({\theta}^m_{\scriptscriptstyle TP,\ell})\} + z\sigma_n^2\textbf{I}_N.
\end{split}
$}
\end{equation}
Here, the $(p,q)^{th}$ element of matrix $\textbf{R}_{\scriptscriptstyle TP,\ell}$ is given by~\cite{mimo_book_correlation_matrix} 
\begin{equation}
\label{eqn7}
\resizebox{0.9\columnwidth}{!}{$
[\textbf{R}_{\scriptscriptstyle TP,\ell}]_{pq} = \rho z\beta_{\scriptscriptstyle TP,\ell}\int e^{-j\frac{2\pi d}{\lambda}(p-q)\cos(\varphi_{\scriptscriptstyle TP,\ell} + \Tilde{\theta})}f_{\Theta}(\Tilde{\theta})d\Tilde{\theta} + z\sigma_n^2\delta_{pq}
$},
\end{equation}
where $\delta_{pq}$ is the Kronecker delta function. Similar to the RSS, $\textbf{R}_{\scriptscriptstyle TP,\ell}$ is estimated by averaging over a large, but finite number of received symbols. The eigenvector $\textbf{u}_1 \in \mathbb{R}^{N \times 1}$ corresponding to the largest eigenvalue of $\textbf{R}_{\scriptscriptstyle TP,\ell}$ span the signal subspace $\textbf{u}_s = [\textbf{u}_1] \in \mathbb{R}^{N \times 1} $, while the eigenvectors $\textbf{u}_2, \textbf{u}_3, ... ,\textbf{u}_N \in \mathbb{R}^{N \times 1}$ corresponding to the $N-1$ smallest eigenvalues span the noise subspace $\textbf{U}_n = [\textbf{u}_2,\textbf{u}_3,\ \dots \ ,\textbf{u}_N] \in \mathbb{R}^{N \times (N-1)}$. The orthogonality of the noise subspace and the
signal subspace implies that the matrix product ${\textbf{a}^H({\theta})\textbf{U}_n\textbf{U}^H_n\textbf{a}({\theta})}$ attains its minimum when the variable $\theta$ is equal to the true AOA $\varphi_{\scriptscriptstyle TP,\ell}^{\circ}$. Hence, the MUSIC pseudospectrum function defined as \cite{180_degree_ambiguity}
\begin{equation} 
\label{eqn8}
\mathrsfso{P}_{\scriptscriptstyle MUSIC}(\theta) = \frac{1}{{\textbf{a}^H({\theta})\textbf{U}_n\textbf{U}^H_n\textbf{a}({\theta})}},
\end{equation}
produces a sharp peak when $\theta = {\varphi}_{\scriptscriptstyle TP,\ell}^{\circ}$. Hence, the estimated AOA $\widehat{\varphi}_{\scriptscriptstyle TP,\ell}^{\circ}$ of the UE is the angle corresponding to this peak in the MUSIC pseudospectrum.

\vspace{-1mm}
\subsection{Positioning Strategy Overview}
Using the available $L$ AOA and $L$ RSS vectors, the CPU constructs a combined AOA fingerprint matrix $\boldsymbol{\Phi} = [\boldsymbol{\Phi}_{1}, \boldsymbol{\Phi}_{2}, \  \dots\  , \boldsymbol{\Phi}_{L}] \in \mathbb{R}^{K\times{L}}$ and a combined RSS fingerprint matrix $\boldsymbol{\xi} = [\boldsymbol{\xi}_{1}, \boldsymbol{\xi}_{2}, \  \dots\  , \boldsymbol{\xi}_{L}] \in \mathbb{R}^{K\times{L}}$. The RSS fingerprint matrix $\boldsymbol{\xi}$, the AOA fingerprint matrix $\boldsymbol{\Phi}$ and the RP position matrix $\boldsymbol{Q}_{\scriptscriptstyle RP}$ collectively form the fingerprint database. This database is used by the CPU to learn a GPR model for each RP coordinate vector in the offline phase, i.e.,
\begin{equation} 
\label{eqn9}
\begin{split}
col_1(\boldsymbol{Q}_{\scriptscriptstyle RP}) = [x_1,x_2,\  \dots\ ,x_K]^\top = f_1(\boldsymbol{\Omega}) + \boldsymbol{\epsilon}_1 \\
col_2(\boldsymbol{Q}_{\scriptscriptstyle RP}) = [y_1,y_2,\  \dots\ ,y_K]^\top = f_2(\boldsymbol{\Omega}) + \boldsymbol{\epsilon}_2
\end{split}
\end{equation}
for $k \in \{1,2,\ \dots \ ,K\}$. Here, $col_i(\cdot)$ indicates the $i^{th}$ column of the matrix and $\boldsymbol{\Omega} = [\boldsymbol{\xi} , \boldsymbol{\Phi}] \in \mathbb{R}^{K\times{2L}}$ is the hybrid fingerprint matrix constructed
by the CPU. The vectors $\boldsymbol{\epsilon}_1,  \boldsymbol{\epsilon}_2 \in \mathbb{R}^{K \times 1}$ represent observation noise arising from measurement inaccuracies inherent in practical implementations of the fingerprinting process. Each component of $\boldsymbol{\epsilon}_1,  \boldsymbol{\epsilon}_2$ is modeled as a zero-mean Gaussian random variable with variances $\sigma_{\epsilon_1}^2$ and $\sigma_{\epsilon_2}^2$, respectively.

Using the learned functions $f_1(\cdot)$ and $f_2(\cdot)$ from GPR model the CPU estimates $\boldsymbol{q}_{\scriptscriptstyle TP}$ as $\overline{\boldsymbol{q}}_{est} = (\overline{x}_{est},\overline{y}_{est}) \in \mathbb{R}^{1\times2}$ in the online phase, after the test point RSS vector $\boldsymbol{\xi}_{\scriptscriptstyle TP} = [\widehat{\xi}_{\scriptscriptstyle TP,1}^{\scriptscriptstyle dB}, \widehat{\xi}_{\scriptscriptstyle TP,2}^{\scriptscriptstyle dB},\ \dots \ , \widehat{\xi}_{\scriptscriptstyle TP,L}^{\scriptscriptstyle dB}] \in \mathbb{R}^{1\times{L}}$ and the test point AOA vector $\boldsymbol{\Phi}_{\scriptscriptstyle TP} = [\widehat{\varphi}_{\scriptscriptstyle TP,1}^{\circ}, \widehat{\varphi}_{\scriptscriptstyle TP,2}^{\circ},\ \dots \ , \widehat{\varphi}_{\scriptscriptstyle TP,L}^{\circ}] \in \mathbb{R}^{1\times{L}}$ is computed by the CPU. The CPU transmits $\overline{\boldsymbol{q}}_{est}$ to the UE via the APs, which is the final predicted location. The primary objective of the positioning process is to bring the estimated position $\overline{\boldsymbol{q}}_{est}$ as close as possible to the true position $\boldsymbol{q}_{\scriptscriptstyle TP}$, effectively minimizing the positioning error $q_{err}$, given by,
\begin{equation}
\label{eqn10}
\resizebox{0.89\columnwidth}{!}{$
\begin{split}
q_{err} &= \|\boldsymbol{q}_{\scriptscriptstyle TP} - \overline{\boldsymbol{q}}_{est}\| = \sqrt{(x_{\scriptscriptstyle TP}-\overline{x}_{\scriptscriptstyle est})^2+(y_{\scriptscriptstyle TP}-\overline{y}_{\scriptscriptstyle est})^2}.
\end{split}
$}
\end{equation}

\section{Hybrid Fingerprint-Based Positioning \\
Using GPR}
\label{fingerprint_pos_section}
Eq. (\ref{eqn4}) and (\ref{eqn5}) show that the RSS (in dB) of a UE measured at an AP is inversely proportional to the logarithm of the distance between them. Angle information (from $\boldsymbol{\Phi}$) and distance information (from $\boldsymbol{\xi}$) derived from the hybrid fingerprint $\boldsymbol{\Omega}$, serve as fundamental elements for the GPR model to estimate $\boldsymbol{q}_{\scriptscriptstyle TP}$ using the hybrid test point vector $\boldsymbol{\Omega}_{\scriptscriptstyle TP} = [\boldsymbol{\xi}_{\scriptscriptstyle TP},\boldsymbol{\Phi}{\scriptscriptstyle TP}] \in \mathbb{R}^{1 \times 2L}$ in the online phase.

The GPR approach assumes that the functions $f_1(\cdot)$  and $f_2(\cdot)$ follow Gaussian processes with zero mean and user-defined covariance functions $\textit{\textbf{C}}_1$ and $\textit{\textbf{C}}_2$ respectively \cite{fingerprint_savic,gpr_book_mit}. $\textit{\textbf{C}}_1$ and $\textit{\textbf{C}}_2$ model the covariance of the $x$ and $y$ coordinates, respectively, of any two RPs in the system as a function of both their
RSS and AOA vectors. The selection of a particular covariance function is contingent upon the nature of the problem at hand and can warrant a study of its own. In this paper we adopt the popular squared exponential kernel, i.e.,
\begin{equation} 
\label{eqn11}
{k}_i(\boldsymbol{r},\boldsymbol{r'}) = b_i^2\exp\left(-\frac{\|\boldsymbol{r}-\boldsymbol{r'}\|^2}{2\varrho_i}\right).
\end{equation}
For GPR modeling, the vectors $\boldsymbol{r},\boldsymbol{r}'$ $\in \{\boldsymbol{\Omega}_j,\boldsymbol{\Omega}{\scriptscriptstyle TP}\}$. Here $\boldsymbol{\Omega}_j = row_j(\boldsymbol{\Omega})$ represents the $j^{th}$ row of the fingerprint matrix $\boldsymbol{\Omega}$ for $j \in \{ 1,2, \dots, K \}$, and $b_i^2, \varrho_i$ are hyperparameters, for~$i \in~\{ 1,2 \}$. Eq.(\ref{eqn9}) can be represented using the Gaussian process assumption of functions $f_1(\cdot)$ and $f_2(\cdot)$ as 
\begin{equation} 
    col_i(\boldsymbol{Q}_{\scriptscriptstyle RP}) \sim \\ \mathcal{N}\left(\textbf{0}_{K\times1},\bm{\mathcal{K}}_i(\boldsymbol{\Omega},\boldsymbol{\Omega}) + \sigma_{\epsilon_i}^2\textbf{I}_K \right )
\label{eqn12}
\end{equation} 
for $i$ $\in$ $\{1,2\}$. Here
$\begin{bmatrix}
\textit{\textbf{C}}_i
\end{bmatrix}_{mn} = 
\begin{bmatrix}
\bm{\mathcal{K}}_i(\boldsymbol{\Omega},\boldsymbol{\Omega})
\end{bmatrix}_{mn}  = 
{k}_i(\boldsymbol{\Omega}_m,\boldsymbol{\Omega}_{n})$, models the covariance between all pairs of hybrid fingerprint vectors, with $[\cdot]_{mn}$ referring to the $(m, n)^{th}$ element of the matrix. The training process involves learning the hyperparameters $b_i^2, \varrho_i$ and the error variances $\sigma_{\epsilon_i}^2$ from the fingerprint database using maximum-likelihood approach as~\cite{ml_rss_surya}
\begin{equation} 
\label{eqn13}
\overline{\boldsymbol{\Gamma}}_i = \underset{\boldsymbol{\Gamma}_i}{\operatorname{arg\,max}} \Big\{ \log\big(p\left(\text{col}_i(\boldsymbol{Q}_{\scriptscriptstyle RP}) \mid \boldsymbol{\Omega}, \boldsymbol{\Gamma}_i\right)\big) \Big\}, 
\end{equation}
where $\boldsymbol{\Gamma}_i =[b_i^2, \varrho_i$,  $\sigma_{\epsilon_i}^2]^T \in \mathbb{R}^{3 \times 1}$ and $\overline{\boldsymbol{\Gamma}}_i$ represents the optimized vector after learning. The non-convex optimization problem in (\ref{eqn13}) can be solved using gradient ascent methods, such as stochastic gradient or conjugate gradient~\cite{gpr_book_mit,ml_rss_surya}.

The joint Gaussian distributions of the RP position coordinates with the test point position coordinates is given by
\begin{align*}
& \begin{bmatrix}
    col_i(\boldsymbol{Q}_{\scriptscriptstyle RP}) \\
    col_i(\boldsymbol{q}_{\scriptscriptstyle TP})
\end{bmatrix}
\sim \\
&\resizebox{0.89\columnwidth}{!}{$
\mathcal{N}
\left(
\textbf{0}_{(K+1)\times1},
\begin{bmatrix}
    \bm{\mathcal{K}}_i(\boldsymbol{\Omega},\boldsymbol{\Omega}) + \sigma_{\epsilon_i}^2\textbf{I}_K & 
    \bm{\mathcal{K}}_i(\boldsymbol{\Omega},\boldsymbol{\Omega}_{\scriptscriptstyle TP})
    \\
    \bm{\mathcal{K}}_i(\boldsymbol{\Omega}_{\scriptscriptstyle TP},\boldsymbol{\Omega})
    &
    {k}_i(\boldsymbol{\Omega}_{\scriptscriptstyle TP},\boldsymbol{\Omega}_{\scriptscriptstyle TP})
\end{bmatrix}
\right )
$}
\tag{14}
\label{eqn14}
\end{align*}
for $i$ $\in$ $\{1,2\}$. Here $\begin{bmatrix} \bm{\mathcal{K}}_i^\top(\boldsymbol{\Omega}_{\scriptscriptstyle TP},\boldsymbol{\Omega})
\end{bmatrix}_{m1}
 = 
 \begin{bmatrix}
\bm{\mathcal{K}}_i(\boldsymbol{\Omega},\boldsymbol{\Omega}_{\scriptscriptstyle TP})
\end{bmatrix}_{m1}
 =
{k}_i(\boldsymbol{\Omega}_m,\boldsymbol{\Omega}_{\scriptscriptstyle TP}) 
$, for $m \in \{ 1,2, \dots, K \}$, models the covariance between the $K$ training fingerprint vectors and the test vector. Conditioning the joint distribution in (\ref{eqn14}) on the fingerprint database matrices $\boldsymbol{\Omega}$ and $\boldsymbol{Q}_{\scriptscriptstyle RP}$ and the test vector $\boldsymbol{\Omega}_{\scriptscriptstyle TP}$, we obtain normal posterior densities for $\boldsymbol{q}_{\scriptscriptstyle TP}$, i.e.,
\begin{equation} 
\tag{15}
\label{eqn15}
p\left(col_i(\boldsymbol{q}_{\scriptscriptstyle TP})|col_i(\boldsymbol{Q}_{\scriptscriptstyle RP}),\boldsymbol{\Omega},\boldsymbol{\Omega}_{\scriptscriptstyle TP}\right) ~  \sim \mathcal{N}(\mu^{est}_{i},{{v}}^{est}_{i}),
\end{equation}
with the mean and variance
\begin{align*}  
&\mu^{est}_{i}  =  \bm{\mathcal{K}}_i(\boldsymbol{\Omega}_{\scriptscriptstyle TP},\boldsymbol{\Omega})[\bm{\mathcal{K}}_i(\boldsymbol{\Omega},\boldsymbol{\Omega}) + \sigma_{\epsilon_i}^2\textbf{I}_K]^{-1}col_i(\boldsymbol{Q}_{\scriptscriptstyle RP}) \\
\tag{16}
\label{eqn16}
&{v}^{est}_{i} = {k}_i(\boldsymbol{\Omega}_{\scriptscriptstyle TP},\boldsymbol{\Omega}_{\scriptscriptstyle TP})\  - \\
&\qquad \qquad  \bm{\mathcal{K}}_i(\boldsymbol{\Omega}_{\scriptscriptstyle TP},\boldsymbol{\Omega})[\bm{\mathcal{K}}_i(\boldsymbol{\Omega},\boldsymbol{\Omega}) + \sigma_{\epsilon_i}^2\textbf{I}_K]^{-1}\bm{\mathcal{K}}_i(\boldsymbol{\Omega},\boldsymbol{\Omega}_{\scriptscriptstyle TP})
\end{align*} 
for $i$ $\in$ $\{1,2\}$. The mean vector $\overline{\boldsymbol{q}}_{est} = (\overline{x}_{est},\overline{y}_{est}) = (\mu^{est}_{1},\mu^{est}_{2})$ is the MMSE estimate of $\boldsymbol{q}_{\scriptscriptstyle TP} = (x_{\scriptscriptstyle TP} , y_{\scriptscriptstyle TP})$ and the variances $v^{est}_1$ and $v^{est}_2$ model the uncertainty in the estimate of $x_{\scriptscriptstyle TP}$ and $y_{\scriptscriptstyle TP}$ respectively.

The computational complexity of the offline phase involves the estimation of hyperparameters $b_i^2$, $\varrho_i$ and error variances $\sigma_{\epsilon_i}^2$, computation of covariance matrices $\bm{\mathcal{K}}_i(\boldsymbol{\Omega},\boldsymbol{\Omega})$ and  the inversions of $[\bm{\mathcal{K}}_i(\boldsymbol{\Omega},\boldsymbol{\Omega}) + \sigma_{\epsilon_i}^2\textbf{I}_K]$, $i$ $\in$ $\{1,2\}$. These computations are primarily dominated by the inversions of the $(K \times K)$ matrices, which are achievable in most $\mathcal{O}(K^3)$ operations. In the online phase, AOA estimation via the MUSIC algorithm incurs a complexity of $\mathcal{O}(N^3)$ \cite{180_degree_ambiguity}, while matrix operations in (\ref{eqn16}) for computing $\overline{\boldsymbol{q}}_{est}$ require $\mathcal{O}(K^2)$ operations \cite{fingerprint_savic}, resulting in a total complexity of $\mathcal{O}(N^3 + K^2)$ per UE. Unlike cellular massive MIMO systems, where $N$ is large, in cell-free systems, $N$ is typically small \cite{cell_free_mimo_book}. Consequently, the overall complexity in the online phase is effectively $\mathcal{O}(K^2)$, which is quadratic in the database size and can be efficiently handled by modern network CPUs.

\section{Simulation Results}
\label{sim_results}
In this section, we evaluate the performance of the hybrid GPR introduced in Section \ref{fingerprint_pos_section}, in comparison to conventional GPR approaches that utilize either AOA or RSS as inputs. We first present the parameters of the cell-free system employed in the simulations and subsequently discuss the regression results.

\subsection{Cell-Free System Parameters}
The APs are deployed randomly in an urban environment of network area 200m$\times$200m. RPs are systematically arranged in a square pattern, forming a grid that spans the entire simulation area, as shown in Fig. \ref{system_model_fig}. Each large-scale fading coefficient is calculated using (\ref{eqn5}) with $p^{0}_{\ell}=-28.8\ dB$ at $d_{\ell}^{0} = 1m$ and $\gamma=3.53$. These parameters are derived from the 3GPP 38.901 urban micro street canyon NLOS path loss model. The shadowing terms from an AP to distinct location points in the network area is correlated as~\cite{ml_rss_surya, cell_free_mimo_book}
\begin{equation}
\tag{17}
\label{eqn17}
\mathbb{E}\{\nu_{m\ell}\nu_{ij}\} = 
\begin{cases} 
\sigma_{\scriptscriptstyle SF}^2 \cdot 2^{-\frac{d_{\scriptscriptstyle mi}}{d_{\scriptscriptstyle corr}}}, & \text{if } \ell = j, \\
    0, & \text{otherwise}.
\end{cases}
\end{equation}
Here, $\nu_{m\ell}$ is the shadowing from AP $\ell$ to location point $m$, $d_{mi}$ is the distance between locations $i$ and $m$ and $d_{corr}$ is the decorrelation distance that is characteristic of the environment. The location points correspond to the RPs and the test point during a positioning exercise.

The scatterers surrounding the UE are modeled as being distributed in accordance with the disk scattering model. For this model, the covariance matrix $\textbf{R}^{ds}_{\scriptscriptstyle TP,\ell}\in \mathbb{R}^{N \times N}$ of the received signal at AP $\ell$, for $z=1$ is given by \cite{disk_scatter_model}
\begin{equation}
\tag{18}
\label{eqn18}
\textbf{R}^{ds}_{\scriptscriptstyle TP,\ell} = \rho\beta_{\scriptscriptstyle TP,\ell}\textbf{G}(\zeta)\odot \textbf{a}(\varphi_{\scriptscriptstyle TP,\ell}^{\circ})\textbf{a}^H(\varphi_{\scriptscriptstyle TP,\ell}^{\circ}) + \sigma_n^2\textbf{I}_N.
\end{equation}
Here $\odot$ denotes the Hadamard product, $\textbf{G}(\zeta) \in \mathbb{R}^{N \times N}$ is the matrix of scaling factors with elements $\begin{bmatrix}
\textbf{G}(\zeta)
\end{bmatrix}_{mn} = [J_0((m-n)\zeta) + J_2((m-n)\zeta)]$,  $\zeta = \frac{2\pi d}{\lambda}\Delta\sin(\varphi_{\scriptscriptstyle TP,\ell}^{\circ})$, $\Delta$ is the single side angular spread of the multipath components impinging the AP antenna array according to the disk scattering model and $J_k$ is the Bessel function of the first kind and order $k$. Eq (\ref{eqn18}) is valid for small angular spreads such that $\sin(\Delta) \approx \Delta$.

Given that the AOA is intended to be measured through geometrical methods during the offline stage, we introduce a nominal error of $\mathcal{N}(0,4)$ degree to the true azimuth AOA of every AP. This serves to model the inherent measurement errors encountered in practical applications. Also, we disregard the 180-degree ambiguity inherent to ULAs, assuming it can be resolved, for example, by analyzing the ambiguous angle pairs obtained from the MUSIC algorithms of multiple APs or through other spatial processing techniques \cite{180_degree_ambiguity}. Table \ref{tab:table1} provides additional simulation parameters.

\begin{table}[!t]
\caption{Simulation Parameters\label{tab:table1}}
\centering
\begin{tabular}{|c|c|}
\hline
\textbf{Parameter} & \textbf{Value}\\
\hline
Carrier frequency & 2 GHz\\
\hline
Signal bandwidth & 10 MHz\\
\hline
Number of APs (L) & 25 \\
\hline
Antenna array spacing (d) & $0.5\lambda$ \\
\hline
Height of an AP & 10 m \\
\hline
Height of the UE & 1.5 m \\
\hline
UE transmit power ($\rho$) & 100 mW \\
\hline
Receiver noise figure & 8 dB \\
\hline
Noise power ($\sigma_n^2$) & -96 dBm \\
\hline
Standard deviation of shadow fading ($\sigma_{\scriptscriptstyle SF}$) & 8 dB \\
\hline
Decorrelation distance ($d_{\scriptscriptstyle corr}$) & 13 m \\
\hline
Number of distinct setups, each with randomly placed APs & 100\\
\hline
Number of randomly placed test points per setup & 1000\\
\hline
Received signal samples used for estimating RSS and  $\textbf{R}^{ds}_{\scriptscriptstyle TP,\ell}$ & 200 \\
\hline

Single side angle spread ($\Delta$)& 10\textdegree \\
\hline
\end{tabular}
\vspace{-4mm}
\end{table}
\vspace{-2mm}
\subsection{Simulation Results and Discussions}

\begin{figure}
\centering
\captionsetup{justification=centering}
\includegraphics[width=2.6 in]{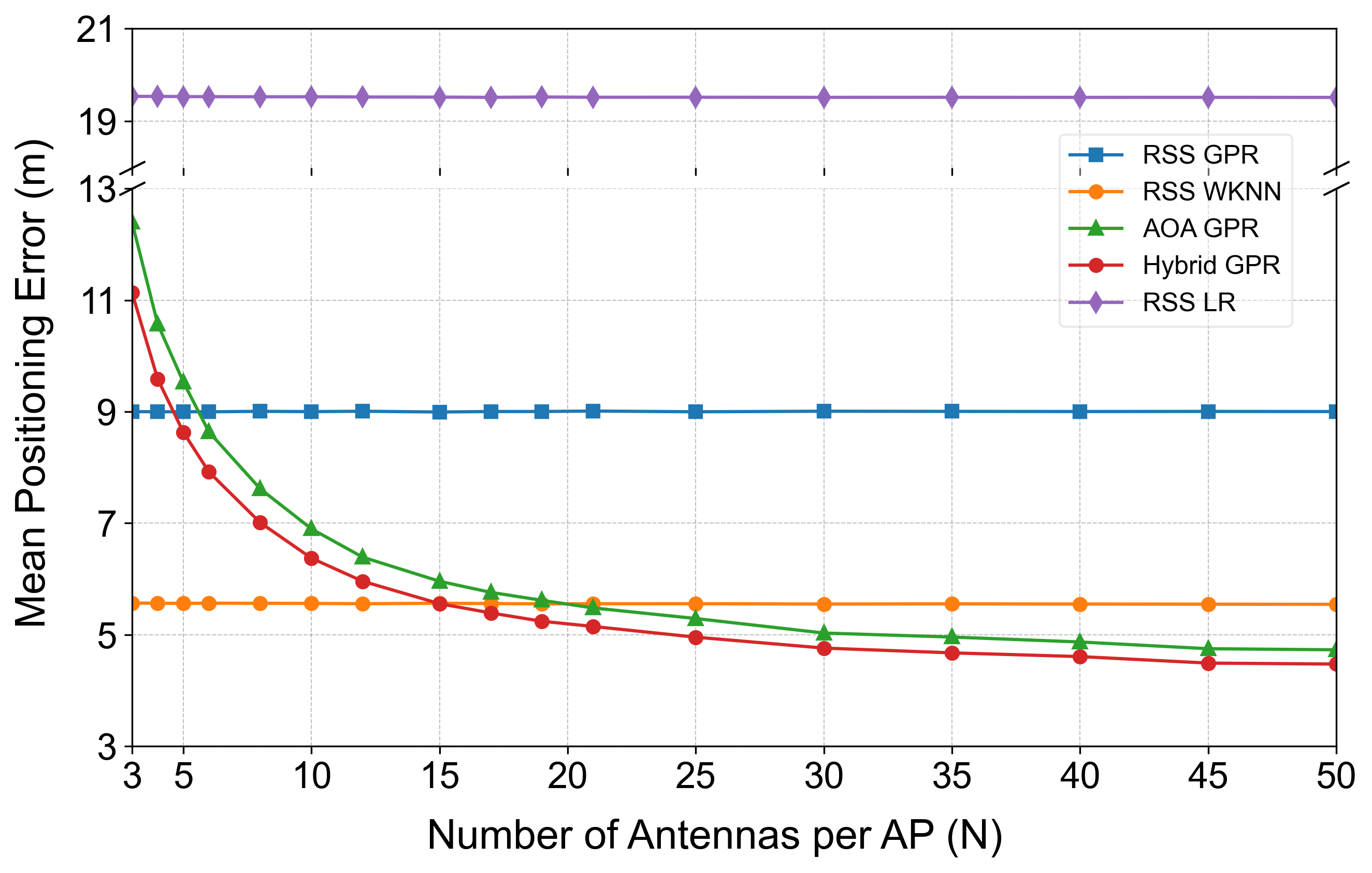}
\caption{Positioning performance for different number of AP antennas ($N$) with $L=25$ and $K=225$ }
\label{N_vs_Err}
\vspace{-3.8mm}
\end{figure}

Fig. \ref{N_vs_Err} shows the average positioning error (across 100 setups with 1000 test points per setup) corresponding to various values of $N$ for different positioning methods. In this context, ``\textit{AOA GPR}'' and ``\textit{RSS GPR}'' refer to regression models that use only AOA and RSS fingerprints, respectively. Specifically, the fingerprint matrix is constructed as $\boldsymbol{\Omega} = [\boldsymbol{\Phi}] \in \mathbb{R}^{K\times{L}}$ and the test vector as $\boldsymbol{\Omega}_{\scriptscriptstyle TP} = [\boldsymbol{\Phi}_{\scriptscriptstyle TP}]  \in \mathbb{R}^{1\times{L}}$ for ``\textit{AOA GPR}''. Similarly,  $\boldsymbol{\Omega} = [\boldsymbol{\xi}] \in \mathbb{R}^{K\times{L}}$ and $\boldsymbol{\Omega}_{\scriptscriptstyle TP} = [\boldsymbol{\xi}_{\scriptscriptstyle TP}] \in \mathbb{R}^{1 \times L}$ for ``\textit{RSS GPR}''. Notably, all equations from section \ref{fingerprint_pos_section} remain applicable to these two regression methodologies. 

Since the performance of the AOA estimation with the MUSIC algorithm improves with increasing number of antennas ($N$), it results in a concurrent decrease in positioning error for both the hybrid and AOA regression methods. Furthermore, the performance of the GPR algorithms can be compared to the WKNN with $k=4$ nearest neighbors and LR algorithms, as detailed in \cite{fog_massive_mimo_ml}. Despite the increase in RSS with $N$, the positioning error for the baseline RSS-based methods remains unaffected, as position information in the RSS vector is solely carried by the large-scale fading coefficient $\beta_{k\ell}$. Notably, RSS-WKNN outperforms hybrid GPR  for fewer antennas per AP, as the latter relies equally on both RSS and AOA inputs. The performance of AOA and hybrid GPRs improves as the quality of MUSIC based AOA estimate improves with N.

\begin{figure}
\centering
\captionsetup{justification=centering}
\includegraphics[width=2.3 in]{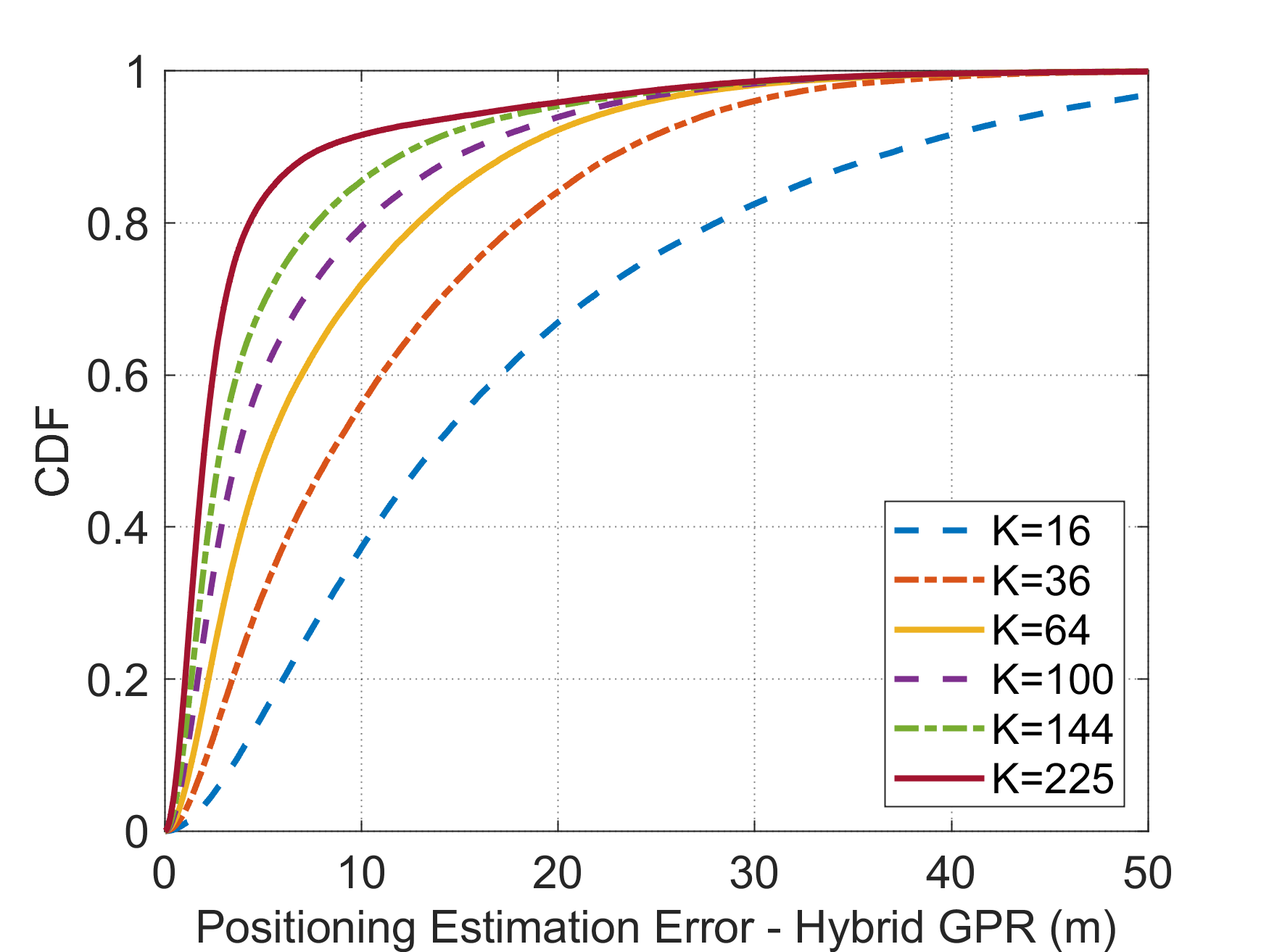}
\caption{CDF of positioning error for hybrid GPR for varying number of RPs ($K$) in the simulation area with $L=25$ and $N = 25$  }
\label{CDF_vs_Err_K}
\vspace{-4mm}
\end{figure}

Fig. \ref{CDF_vs_Err_K} depicts the cumulative distribution function (CDF) of position estimation errors for the hybrid regression across different numbers of RPs covering the simulation area. The curves highlight that as the fingerprint size expands, the positioning performance improves. This enhancement is attributed to the augmented information captured during the offline phase, leading to more accurate training of the regression model. However, this improvement comes at the cost of increased computational complexity in both offline and online stages. Additionally, a practical implementation involves calculating the RSS and AOA values from each of these RPs across the network area, which adds to the burden of fingerprint collection for the system designer. 

\section{Conclusion}
\label{conclusion}
In this paper, a 2D positioning approach using Gaussian process regression was presented. The approach integrated AOA and RSS fingerprints, with a cell-free massive MIMO system serving as the framework for the algorithm. Positioning performance was compared with AOA-only and RSS-only GPRs as well as linear regression and WKNN methods. The results showed that AOA-only and hybrid GPRs achieved lower average positioning errors across multiple setups with sufficient antennas per access point. This study underscores the benefits of integrating AOA and RSS into positioning systems and encourages further exploration of positioning and localization using integrated sensing and communications. 

\vspace{-0.5mm}
\nocite{}
\bibliographystyle{IEEEtranDOI}
\bibliography{IEEEabrv,bibLatex}
\end{document}